\documentclass[a4paper,11pt]{article}
\usepackage{jinstpub} 
\usepackage{lineno}
\usepackage{url}

\usepackage{amsmath}
\usepackage{siunitx}

\title{Design, construction, and operation of a 1-ton Water-based Liquid scintillator detector at Brookhaven National Laboratory}

\author[1,2]{X. Xiang,}
\affiliation[1]{Brookhaven National Laboratory, Upton, NY, USA}
\affiliation[2]{Tsung-Dao Lee Institute, Shanghai Jiao Tong University, Shanghai 201210, China}

\author[1,3,4,*]{G.~Yang, \note[*]{Corresponding author}}
\affiliation[3]{Physics Department, University of California at Berkeley, Berkeley, CA 94720-7300, USA}
\affiliation[4]{Lawrence Berkeley National Laboratory, 1 Cyclotron Road, Berkeley, CA 94720-8153, USA}

\author[1]{S.~Andrade,}
\author[3,4]{M.~Askins,}
\author[1]{D.M.~Asner,}
\author[5]{A.~Baldoni,}
\affiliation[5]{Physics Department, The Pennsylvania State University, State College, PA 16801, USA}
\author[5]{D.F.~Cowen,} 

\author[1]{M.V.~Diwan,}

\author[1]{S.~Gokhale,} 

\author[1, 6]{S.~Hans,} 
\affiliation[6]{Bronx Community College, Bronx, NY 10453, USA}

\author[1]{J.~Jerome,}

\author[7]{G.~Lawley,} 
\affiliation[7]{Physics Department, Stony Brook University, Stony Brook, NY 11794, USA}
\author[1]{S.~Linden,}

\author[3,4]{G.D.~Orebi~Gann,} 

\author[1]{C.~Reyes,} 

\author[1]{R.~Rosero,}

\author[1]{N.~Seberg,}
\author[3,4]{M.~Smiley,} 

\author[1]{N.~Speece-Moyer,}

\author[1]{B.~Walsh,}
\author[8]{J.J.~Wang} 
\affiliation[8]{Department of Physics and Astronomy, University of Alabama, Tuscaloosa, AL 35487, USA}
\author[9]{M.~Wilking,}
\affiliation[9]{School of Physics and Astronomy, University of Minnesota, Minneapolis, MN  55455, USA}
\author[1]{M.~Yeh,}

\abstract{

Water-based liquid scintillators (WbLS) are a new class of detector materials
that provide efficient and tunable detection of both Cherenkov and scintillation light. A massive WbLS neutrino detector with suitable photosensor coverage for low intensity light detection could therefore reconstruct the momentum of an energetic charged  particle and also have enhanced low-energy sensitivity.   
These materials are also better suited for metal doping broadening the potential scientific utility. 
We recently constructed and commissioned a 1-ton WbLS detector with good photosensor coverage and a capable data acquisition and calibration system. We intend to use this flexible detector system as a testbed for WbLS R\&D.  In this paper we give an overview of the  1-ton system and provide some early   results. 
}


\begin{document}

\maketitle

\flushbottom

\section{Introduction}
\label{sec:intro}

A new generation of photon based neutrino detectors is being explored. These are designed to provide exceptional sensitivity to a broad range of neutrino physics, such as long-baseline neutrino oscillations, diffuse supernovae neutrinos, solar neutrinos, and other low-energy neutrino phenomena~\cite{Klein:2022tqr}. A key innovation in this pursuit is the hybrid detection technique, which employs both Cherenkov and scintillation light simultaneously, resulting in excellent event imaging and energy resolution  that significantly improve background rejection, and provide excellent sensitivity to a broad range of physical phenomena.

The hybrid technique has been demonstrated in the Borexino detector, filled with Pseudocumene (PC) for MeV scale neutrino events, and the SNO+ detector half-filled with linear alkyl benzene (LAB) with an admixture of 2,5-Diphenyloxazole (PPO) for similar low energy events~\cite{BOREXINO:2021xzc, SNO:2023cnz}. 
By measuring the time of the earliest detected photons, and by full event-by-event reconstruction, respectively, the Borexino and the SNO+ collaborations have  demonstrated directional sensitivity in large-scale liquid scintillator detectors.
Having established the feasibility of this technique, it is important to optimize and scale it up to multi-kiloton size detectors to obtain next-generation sensitivity.  

The  development of water-based liquid scintillators (WbLS) at Brookhaven National Laboratory (BNL) provides a cost-effective and environmentally suitable solution for realizing kiloton-scale hybrid detectors \cite{YEH201151, ANNIE:2023yny}. WbLS is created by encapsulating Liquid Scintillator (LS) within nanometer-scale micelles, using surfactant chemistry similar to that found in detergents. Each surfactant molecule features a hydrophilic head and a hydrophobic tail, which collectively envelop the oil-like LS droplets to form micelles. The micelles allow LS to be dissoluble in a non-miscible liquid such as water, homogeneously, and the resulting WbLS mixture can remain stable for years across a wide range of temperatures and pH levels.

Aside from affordability and environmental suitability, WbLS offers three main features that make for flexible detector designs. First, WbLS features a tunable scintillation light yield and time profile to meet application-specific needs. With
a relatively long attenuation length, WbLS allows experiments to custom-design large-scale detectors that optimize signal quality and physics reach. Second, WbLS should excel in the realm of particle identification (PID) through its capability for Cherenkov/Scintillation separation (C/S), a critical feature for background suppression. Third, WbLS can be loaded with a variety of metals, such as gadolinium or lithium, to enhance its utility for neutron capture events, making it a versatile choice~\cite{buck_2016, Theia:2019non}.

\begin{figure}[t]
    \resizebox{\textwidth}{!}{%
    \includegraphics[height=4cm]{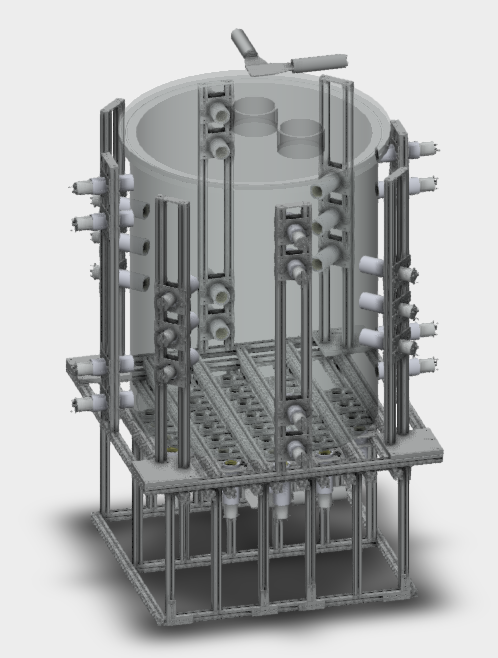}
    \includegraphics[height=4cm]{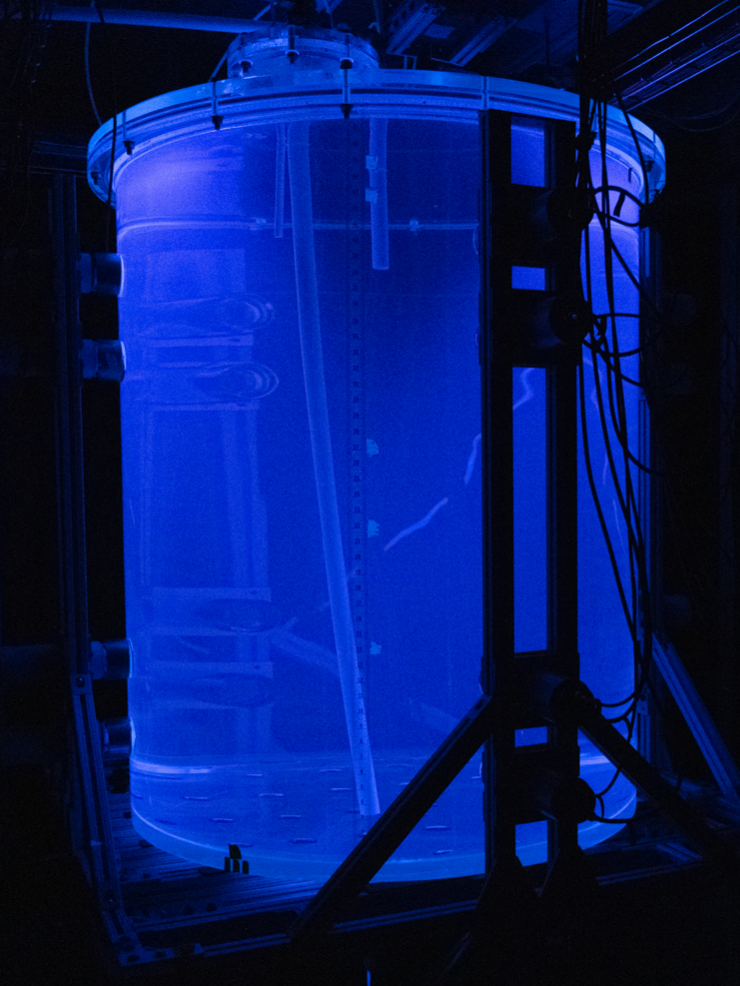} 
    }
    \caption{Left: Illustration of the detector layout. In total, 30 PMTs are on the bottom and 28 on the side. Two muon paddles on the top and two on the bottom. Right: The same 1-ton WbLS liquid is illuminated by two UV lamps positioned at the top, causing it to scintillate blue light.}
    \label{fig:detector}
\end{figure}

Several current and proposed particle and nuclear physics  experiments, such as \textsc{Eos} \cite{Anderson:2022lbb}, BUTTON, 
ANNIE \cite{ANNIE:2015inw}, LiquidO \cite{LiquidO:2019mxd}, and \textsc{Theia} \cite{Theia:2019non} plan to adopt WbLS as a medium to achieve hybrid event detection. Nevertheless, the scalability of WbLS to large scale applications remains an open question. 
In this paper, we show  the design, construction, and operation of a 1-ton WbLS detector surrounded by PMTs as a testbed,  shown in Fig.~\ref{fig:detector}, aiming to establish WbLS as a versatile platform for various high-energy physics applications. The new system we describe below is a considerable upgrade over the previous test with the same acrylic vessel \cite{zhao2024}. It includes a novel circulation and filtration system, much larger photo-coverage, suitable data acquisition, and calibration systems. Most importantly, the new system deploys a newer formulation of the liquid that reduces scattering, and allows us to load the liquid with metals such as Gd. In this paper we provide an overview of these developments since the previous publications~\cite{YEH201151, Bignell:2015msa, buck_2016, diwantipp}.







\section{Optical Detector}
\label{sec:detetcor}

\begin{figure}
    \centering
    \includegraphics[width=0.8\textwidth]{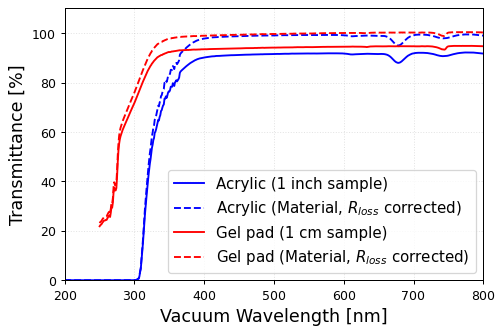}
    \caption{Optical transmission measurements of a Nakano-UVT Acrylic sample (solid blue) and a RTV-6136 silicone optical gel pad sample (solid red) in air. The measured data was compared to ND filters with known transmittance and absorbance, which carries $\pm 2\%$ uncertainty according to data sheet. The dashed curves are the calculated material transmittance after accounting for the Fresnel reflection loss ($R_{loss}$) on both surfaces. The $R_{loss}$ is computed using wavelength-dependent material refractive index at a normal incidence. Both acrylic and gel pad show excellent transparency in the visible spectrum. 
    }
    \label{fig:transmission}
\end{figure}

The 1-ton WbLS testbed detector is composed of  a cylindrical tank made of Nakano-UVT ~\cite{Nakano} acrylic with an outer diameter of 1150 mm and a height of 1275 mm. The tank wall is 25.4 mm thick. When fully filled, the tank holds a volume equivalent to about 1 ton of water. The tank is designed to be stable with excellent impact resilience. The tank has excellent optical clarity in the visible spectrum, with transparency exceeding 95\% down to wavelengths of 380 nm as shown in Fig. ~\ref{fig:transmission}.
The lid of the tank is equipped with two $\sim100 $ mm tall chimneys through which various tank inputs and outputs are arranged including 
circulation of the detector fluid, openings for level sensors,  calibration port, and nitrogen gas for a  blanket above the liquid. This nitrogen layer (kept at a slight pressure) prevents the intrusion of oxygen and carbon dioxide into the liquid. 


\begin{figure}
    \centering
    \includegraphics[width=0.45\textwidth]
  {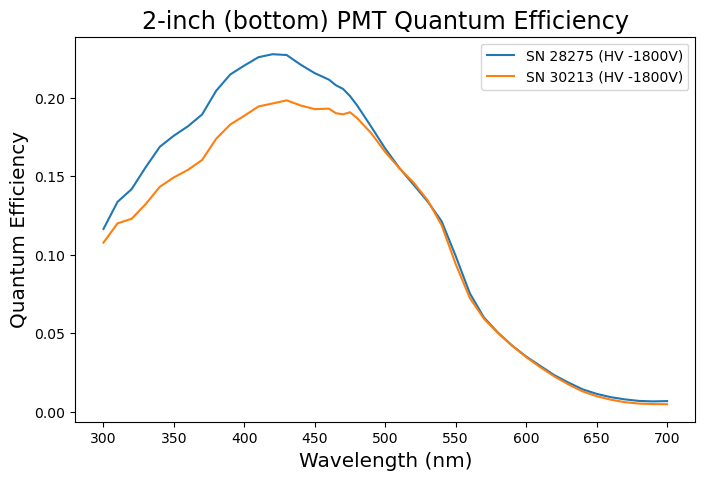}
    \includegraphics[width=0.45\textwidth]
    {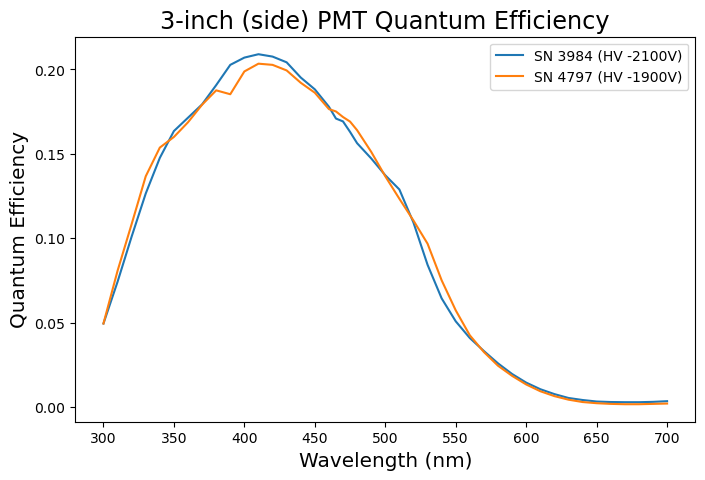}
    \caption{Measurement of
    quantum efficiency for   ET-enterprise 9954B 2-inch (Left) and 9821KB 3-inch (Right) series photo-multipliers. Two PMTs for each version were measured at the center of the photo-cathode. 
    The mean of these values for each PMT type was used in the simulation.  
    The measurement is shown  for $>300$ nm since the acrylic tanks cuts off sensitivity below 300 nm.  }
    \label{fig:qe}
\end{figure}

The tank is positioned within a frame constructed from the 80/20 aluminum building system, which also simplifies the placement of photomultiplier tubes (PMTs) around it, as illustrated in Fig. \ref{fig:detector}. Light from the detector is captured by 58 flat-faced ET-enterprise PMTs situated outside the acrylic tank: 30 are 2-inch 9954B series models located beneath the tank, and 28 are 3-inch 9821B series models positioned along the tank's side. 
The  number of PMTs corresponds approximately to 7.1\% coverage of the bottom area, and 3.0\% coverage of the side. The earlier part of the data, described here,  was with 16 side PMTs corresponding to 1.7\% of the side area. 
No PMTs were placed at the top of the tank viewing downwards, and so the total coverage over all solid angle was 3.2\% (2.3\%) for the later (earlier) part of the data.  
These PMTs are optically coupled to the acrylic tank using silicone gel pads (below referred to as "cookies"). The bottom cookies are 2-inch diameter disks with a constant thickness of 6.35 mm (1/4 inch), while the side custom made cookies have a thickness ranging from 3 mm to 5 mm. The side cookies are specifically designed so their inner surface matches the curvature of the acrylic tank. Made from the same RTV material used in the IceCube experiment~\cite{momentive}, these cookies offer excellent optical transparency down to 300 nm as shown in Fig. ~\ref{fig:transmission}. 
To ensure proper optical coupling between the acrylic tank and the PMTs, tension is applied to the PMT bases using springs or plastic ties.  The PMT quantum efficiency was also measured with a calibrated photo-spectrometer for four PMTs to provide input for the simulation of the apparatus (shown in Fig. \ref{fig:qe}).   

The entire detector setup is housed in a lightproof dark room constructed from aluminum panels that are darkened inside with black PPV sheets to prevent reflections. The room measures $3.5 \times 3.5 \times 2.5$ meters, providing ample space for personnel to enter for construction and maintenance purposes. 

\section{WbLS production and circulation system}
\label{sec:organic} 

\begin{figure}[t]
    \centering
    \includegraphics[width=\textwidth]{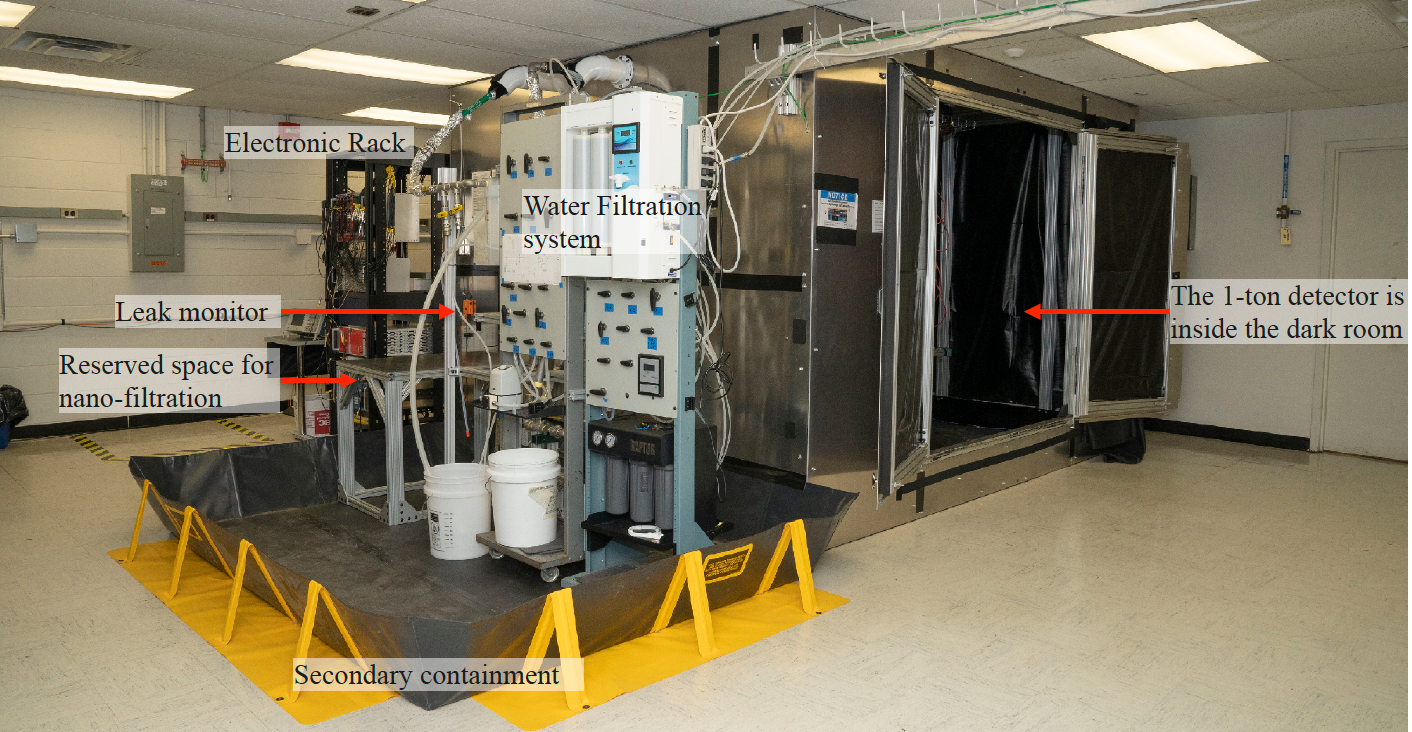}
    \caption{A wide angle view of the 1-ton testbed facility at BNL.}
    \label{fig:full_view}
\end{figure}

The circulation system, pictured  in Fig. \ref{fig:full_view}, consists of a circulation pump, a water filtration system,resistivity sensors, and an injection port for pre-mixed liquid scintillator (LS). All inlet and outlet of the flow pipes are made of PTFE to ensure chemical resistance and purity. Integrated pH meters, flow meters, and thermometers are also part of the system for comprehensive monitoring. 
The water purification and deionization system is capable of achieving 18 M$\Omega$-cm high resistivity  water. In the WbLS phase  the purification system is used to ensure high purity for WbLS without deionization.  
The high-purity WbLS has been able to circulate through the system for many weeks  without requiring further processing, maintaining optical properties consistent with laboratory samples that are measured  in a benchtop system  to have scattering lengths of several 
tens of meters. In compliance with laboratory safety guidelines, a leak monitor was installed and all components and the detector are housed within secondary containment structures.

To create a water-like WbLS, it is essential to introduce a small fraction of liquid scintillator in a water medium. However, due to the polarity differences between organics and water, combining them homogeneously is challenging and, in most cases, takes many days of mixing for a liter-sized sample. To overcome this, a refined formula allowing fast and homogeneous emulsification of LS in water has been developed. This formulation reduces the liquid-liquid interfacial tension, resulting in a stable WbLS that is ideal for a wide range of applications. The new WbLS formulation used in the  1-ton system consisted of a mix of modified surfactants, a fluor of 2,5-diphenyloxazole (PPO), and a DIN-based (di-isopropylnaphthalene) liquid scintillator, all commercially available. All the raw components were purified by in-house thin-film vacuum distillation and recrystallization.

One of the main goals of the 1-ton testbed is proving the feasibility of a scaled up WbLS deployment. Conventional liquid scintillator detectors require large scintillator targets produced at a scaled up manufacturing facility, transferred to the location of the detector vessel, and then deployed into the vessel with various controlled measures. This traditional approach is labor and time consuming and demands a large space to accommodate essential facilities. For our current test,  a newly developed in-situ WbLS mixing technique largely simplifies the process by manufacturing the liquid and filling the detector simultaneously. The detector vessel was first filled with pure water. A small quantity of organics was then sequentially injected into the vessel allowing WbLS mixing in-situ through the existing circulation system.

\begin{figure}[t]
    \centering
    \includegraphics[width=0.9\textwidth]{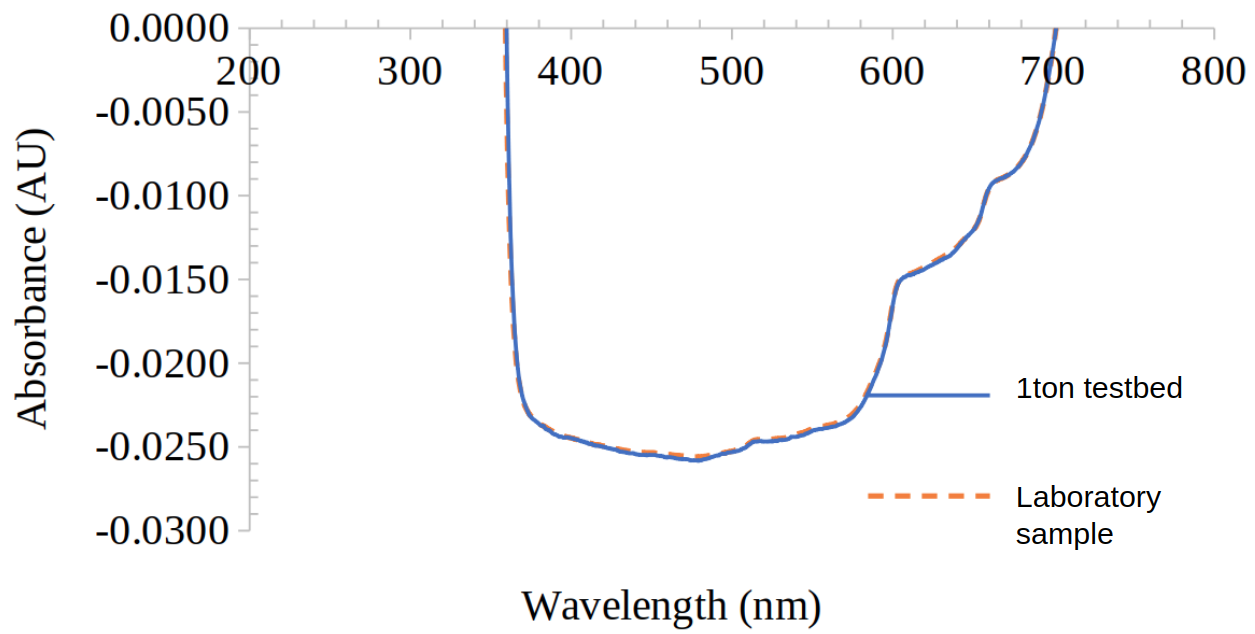}
    \caption{The relative absorbance spectra of WbLS from the 1-ton system compared to laboratory samples 
    }
    \label{fig:uv}
\end{figure}

\begin{figure}[t]
    \centering
    \includegraphics[width=0.9\textwidth]{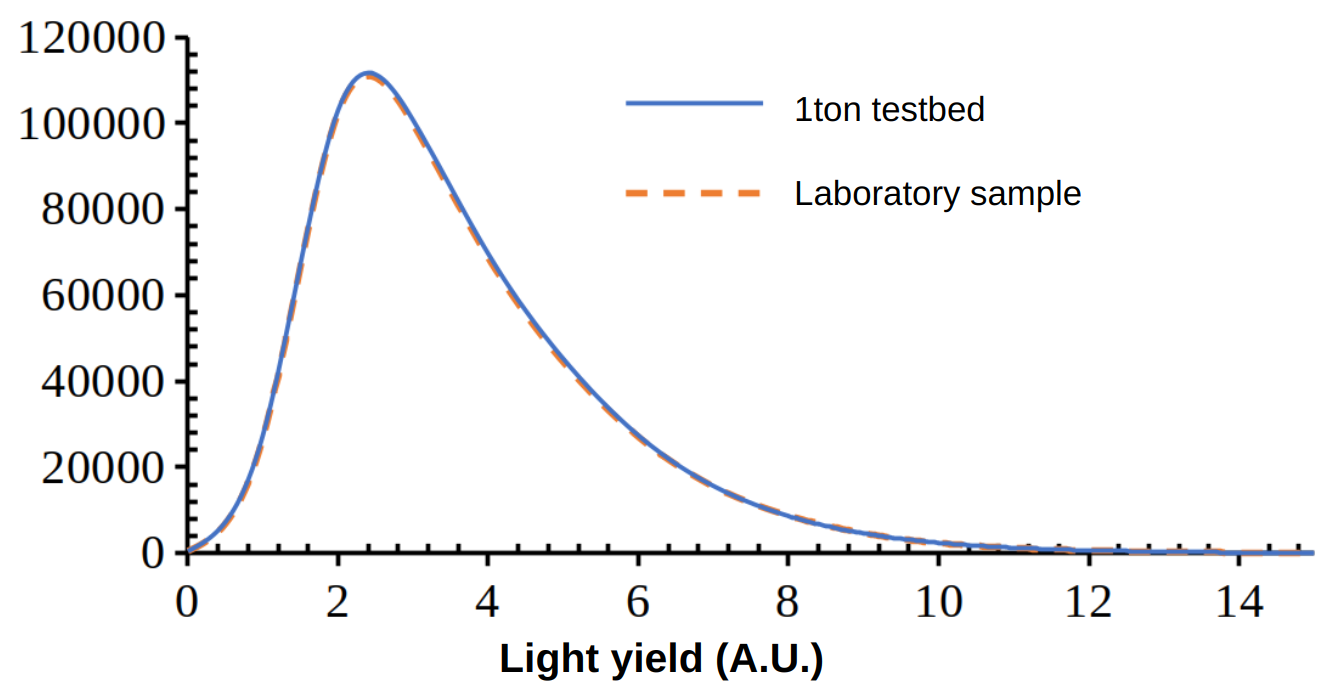}
    \caption{The relative light yield of WbLS from the 1-ton system compared to laboratory samples. }
    \label{fig:lyy}
\end{figure}

Ten kilograms of purified organic materials (1\% of 1-ton) were prepared in stages in a controlled environment. After the completion of organics preparation, the product was filtered using a 0.2-micron polypropylene filter and stored in an injection bottle.
The organics were injected sequentially using a Teflon-based KNF pump at $\sim$1 LPM into the 1-ton testbed \cite{knf}. The concentration of LS in water was precisely measured to be 1\% by mass. 
The WbLS homogeneity was observed 20 mins after injection of organics (section~\ref{injection}). This demonstrated the success of in-situ sequential mixing scheme as shown in Figs.~\ref{fig:uv} and \ref{fig:lyy}. The performance, in terms of optical transmission and light-yield, of WbLS samples from  the  1-ton testbed was measured and compared against a preparation of separate laboratory samples. The tests were done on bench top instruments in identical ways. 
The 1-ton samples and the laboratory-prepared samples show excellent agreement indicating the scalability from a preparation at liter scale  to one-ton scale. The performance and stability were monitored since deployment for several weeks and found to be stable as observed in the benchtop measurements. 

The relative optical absorbance of samples of  WbLS from laboratory preparations and from  the 1-ton system was measured using a Shimadzu UV-Vis spectro-photometer  shown in Fig.~\ref{fig:uv}. The formulation used in the 1-ton system has a scattering length at wavelength of  450 nm higher than 30 meters~\cite{chess}. The relative light yield was measured using a coincidence scintillation counter (Beckman LS6500) shown in Fig.~\ref{fig:lyy}. The counter is equipped with a $^{137}$Cs gamma source, two photomultipliers tubes in coincidence mode, and a multichannel analyzer. This is a relative measurement indicating that the 1-ton and the laboratory samples were identical.  
The absolute light yield of the same material has been measured independently using the CHESS apparatus~\cite{Caravaca:2016ryf} to be approximately 100 photons per MeV. A paper is under preparation to describe this measurement, along with other properties of this material such as the intrinsic scintillation time profile~\cite{yeh2024gd}.



\section{Data Acquisition}
\label{sec:daq}

The data acquisition (DAQ) system has five components: signal readout, multi-ADC trigger synchronization, high voltage (HV) supply, trigger system, and DAQ software.
The signals from all photomultiplier tubes (PMTs) are conveyed directly to CAEN V1730S ADC boards using  cables of equal length to ensure synchronous signal arrival times. Each V1730S digitizer has a sampling frequency of 500 MHz. Our setup comprises four V1730S boards, each equipped with 16 channels. To synchronize the readout ADCs, the clocks of the four V1730S boards are daisy-chained together. Clock phase offsets relative to the master board (the first in the series) are calibrated using PLL files obtained from CAEN, achieving inter-board phase alignment within 150 ps.

For optimal trigger time synchronization, we evaluated both the FAN-OUT option and a daisy-chain setup using a fast square pulser. While the FAN-OUT option mostly synchronized triggers across the V1730S boards, it did exhibit jitter approximately every $\sim 10^4$
pulses. In contrast, although the daisy-chain setup introduced a consistent trigger time delay between consecutive boards, measured to be 48 ns, no jitter was observed between boards over the course of a million test pulses. For these reasons, we opted for the daisy-chain setup and adjusted for the 48 ns delay during offline data processing.
  
The high voltage  for the PMTs is supplied by multiple CAEN A7236N modules, connected via SHV cables. The output current is continuously monitored and logged by GECO2020 software to ensure long-term stability. During this monitoring, we were able to detect several periods of unstable HV behavior, and we have eliminated the corresponding data from the final selection. To further ensure signal quality, the DAQ rack is carefully grounded. We carried out extensive noise hunting campaigns and have reduced the baseline standard deviation to below 0.3 mV for a 2 Vpp (peak to peak)  dynamic range and 0.1 mV for a 0.5 Vpp dynamic range. The 2 and 0.5 Vpp ranges are the two options offered on the V1730S boards and the 2 Vpp was chosen.

The trigger system is designed to record cosmic muons entering the tank. Two rectangular plastic scintillator paddles, each measuring approximately $10 {\rm cm} \times 12 {\rm cm}$ and read out by a single PMT, are positioned atop the tank. The paddles are oriented 90 degrees relative to each other and are separated by approximately 4.5 inches so that the overlapping area is $10 {\rm cm} \times 10 {\rm cm}$. Two lead bricks are placed above the paddles to attenuate  soft radiation induced by cosmic muons, to reduce false triggers. 
When a muon passes through both scintillator paddles and a coincident match occurs, a NIM external trigger signal is generated and sent to the master V1730S board. A second set of same sized  muon paddles are located below the  tank, but these are simply read out in two channels of the V1730s. These paddles allow us to select muons that pass through the tank in a well defined trajectory.  In a later phase of the experiment, we added a calibration source at the center of the detector. This source, described below,  was also viewed by a tiny PMT  inserted into the detector. The signal from this PMT was also included in the trigger system.  

We developed the DAQ software based on the ToolDAQ ~\cite{tooldaq} framework to interface with CAEN digitizers. ToolDAQ is a pure C++ framework with built-in features for logging, execution, variable handling, and serialization. The DAQ software reads data from each board via an optical fiber connected to a CAEN VME bridge and dumps it into binary files. Although this data transmission is limited by the single PCIe speed (<80 MB/s), it's a reliable system, sufficiently facilitating precise analysis and interpretation of detector signals.

In summary, the DAQ system utilizes CAEN V1730S ADC boards for high-frequency signal digitization, achieves synchronization across multiple boards, and relies on ToolDAQ for data management, all while employing a cosmic muon-trigger mechanism to initiate data capture.

\begin{figure}[t]
    \centering
    \includegraphics[width=0.8\textwidth]{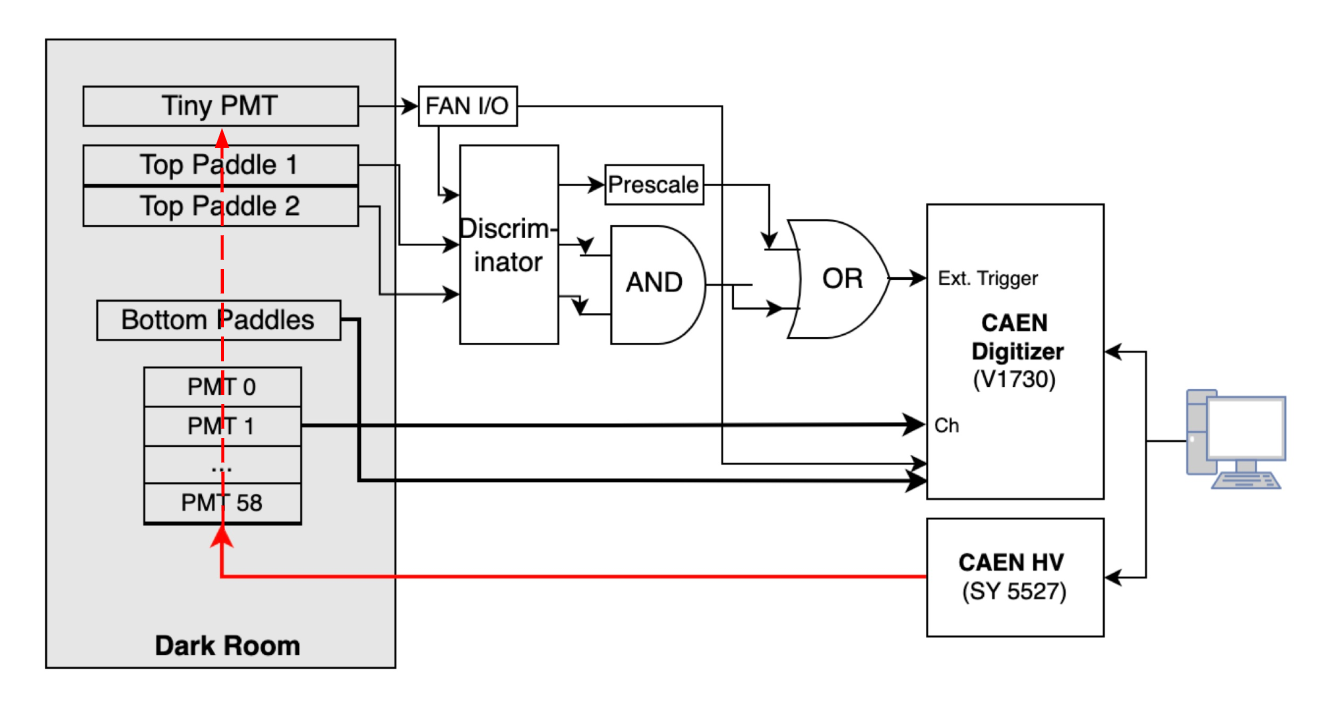}
    \caption{A schematic illustrating the DAQ and trigger system during the muon data taking.  The calibration source (tiny PMT) injected triggers in the same DAQ chain; it was implemented in a later phase of the experiment. A random pulser was often also inserted in the trigger to obtain noise data.  The high voltage bias is shown in red. 
    }
    \label{fig:daq_schematic}
\end{figure}

\section{Detector Calibration}
\label{sec:calibration}


PMT gain ($G$) is a measure of the amplification factor, defined as the number of output electrons at anode ($n_a$), divided by the number of initial photoelectrons (PE) at the cathodes ($n_c$). When an average number of PE produced at the photocathode is less than 1, the single PE can be easily identified in the charge spectrum. 

We successfully conducted PMT gain calibration using a single LED positioned atop the tank. The LED was synchronized with the DAQ system through a pulse generator, which also allowed us to vary the LED intensity by controlling the pulse height. The calibration involved running a series of LED tests at different LED intensity, with the intention to produce the right amount of $n_c$ for all PMTs. Each run  frequency was 1-2 kHz and usually lasted a few minutes. In the data analysis phase, we specifically selected pulses whose peaks occurred within a 40 ns window 
around the expected LED photon arrival time given the LED pulse width is 20 ns. For each PMT, we chose the LED intensity that resulted in an occupancy ranging from 0.01 to 0.04. Here, occupancy is defined as the fraction of triggers where the waveform peak within the 40 ns window exceeded 1.5 mV. 

The Single Photoelectron (SPE) peak was modeled using a simple Gaussian function, from which we extracted the mean to estimate the PMT gain. The Gaussian fit's accuracy was bolstered by meticulous control over pulse timing and occupancy, which effectively minimized dark counts to below $10^{-4}$ for the majority of PMTs. Even though this approach doesn't account for  the skewness in the SPE peak, it serves as a reliable and stable method for monitoring PMT gain. 

We conducted an LED calibration campaign before collecting muon data to standardize the gains across all PMTs. This was achieved in two steps. First, we chose a representative PMT and plotted the relationship between its gain and bias voltage. Second, through an iterative process, we fine-tuned the high-voltage settings for each PMT until all gains were within 20\% of nominal. Throughout the muon data collection period, we monitored the PMT gains daily to weekly to ensure the stability of both the PMTs and the high-voltage supply. 
We also standardized the gains to $6\times 10^6$ to prevent potential saturation from large signals throughout the data taking. The obtained mean gain was used as a conversion factor to convert charge to photo-electrons for each PMT. 

\begin{figure}[t]
    \centering
    \includegraphics[width=0.9\textwidth]{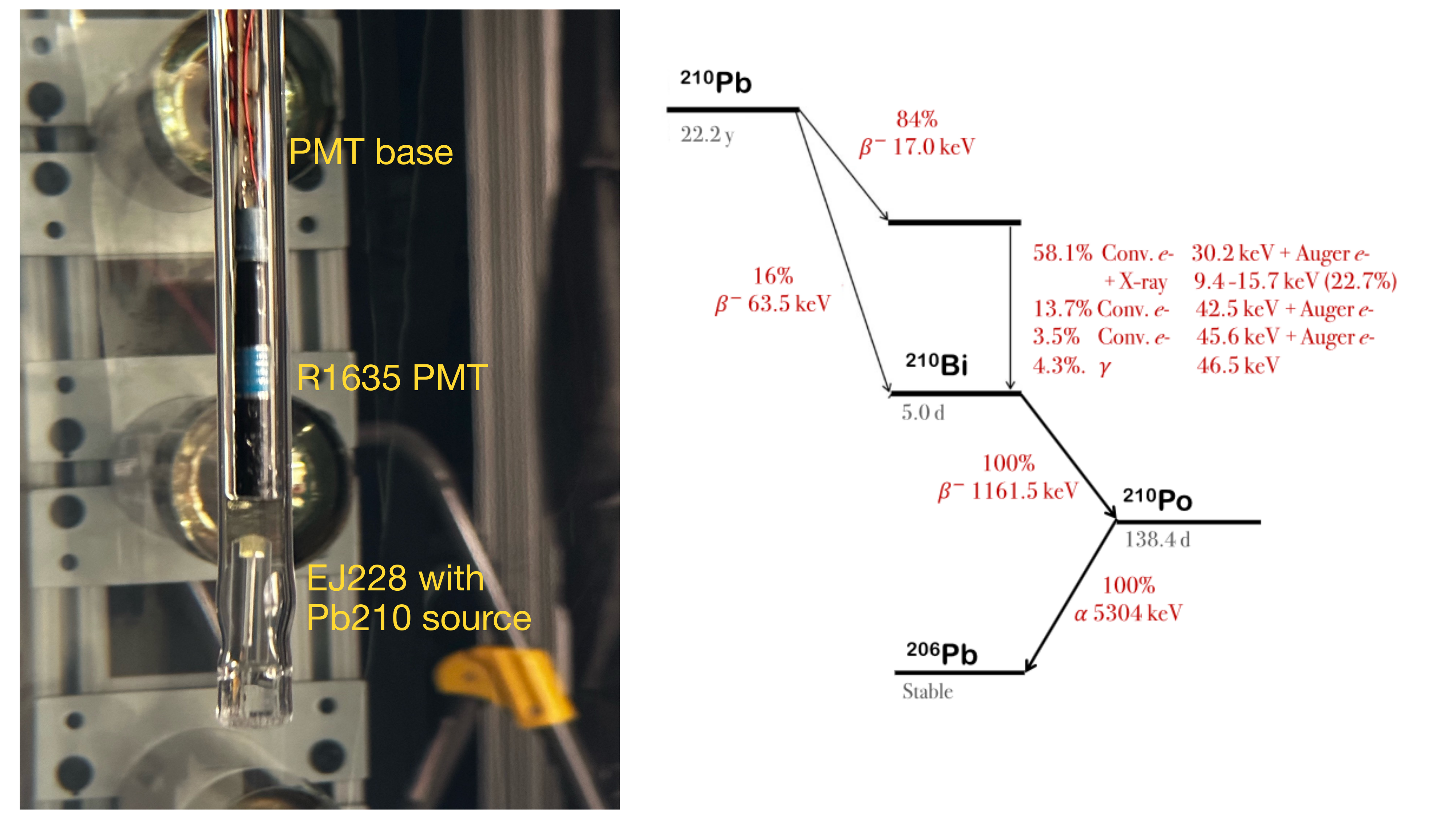}
    \caption{(Left) Photograph of the calibration source as deployed inside the detector.  The source was contained inside a EJ228 plastic scintillator cylinder. The assembly, source, EJ228, a small PMT for tagging the source, and electronics, were inserted in a 10 mm ID acrylic tube which was inserted into the detector so that the source was approximately at the center of the detector. (Right) The decay scheme for $^{210}Pb$ that ends with an alpha decay  from $^{210}Po$. The scintillation light is produced mainly by the 1161 keV beta and the 5305 keV alpha in the plastic scintillator
    ~\cite{Alenkov:2021asw}.  }
    \label{alpha_source}
\end{figure}

A low intensity (0.01 microcurie) radioactive  source was designed for detector calibration for later phases of data-taking (Fig. \ref{alpha_source}). The source consisted of 
$^{210}$Pb deposited on a needle which was placed inside an EJ-228 plastic scintillator~\cite{eljen}. The plastic scintillator was a 5-mm diameter $\times$ 25-mm length cylinder; a small hole was drilled along the axis of the cylinder, the source needle was placed inside and sealed by epoxy. The source was viewed by a small 10 mm diameter PMT (Hamamastu R1635) with a specially designed base which allowed the entire assembly, source, PMT, base, and cables, to be inserted in a clear acrylic tube. The plastic scintillator was coupled to the PMT with GE-RTV6136-D1 2 part silicon compound~\cite{momentive}. The acrylic tube was sealed at the bottom end which held the source, and the entire assembly was rigorously tested against water leakage.    

The $^{210}$Pb decay scheme is shown in Fig. \ref{alpha_source}; the radiation of most interest are the 1161 keV $\beta$ (CSDA range $\sim$5 mm) and the 5305 keV $\alpha$ (CSDA range $\sim$40$\mu$m). Both of these deposit their energies in the EJ-228 plastic and give rise to a  spectrum with an approximate yield of 10k photons per MeV for minimum ionizing $\beta$'s and  $\sim$1k photons per MeV for the $\alpha$'s due to the saturation in the plastic. 
Contributions to the trigger rate and light yield from the other $\beta$s and the $\gamma$ are very small.  Some of this light will cause the small R1635 PMT to trigger and provide a start signal for acquiring the   calibration data. 
The threshold is set to obtain a trigger rate of 20 Hz, which we prescaled down to $\sim$1 Hz for sufficient data rate.
Most of the light from the scintillator escapes the plastic and the acrylic tube and shines isotropically  in the detector.   

\begin{figure}[t]
    \centering
    \includegraphics[width=0.8\textwidth]{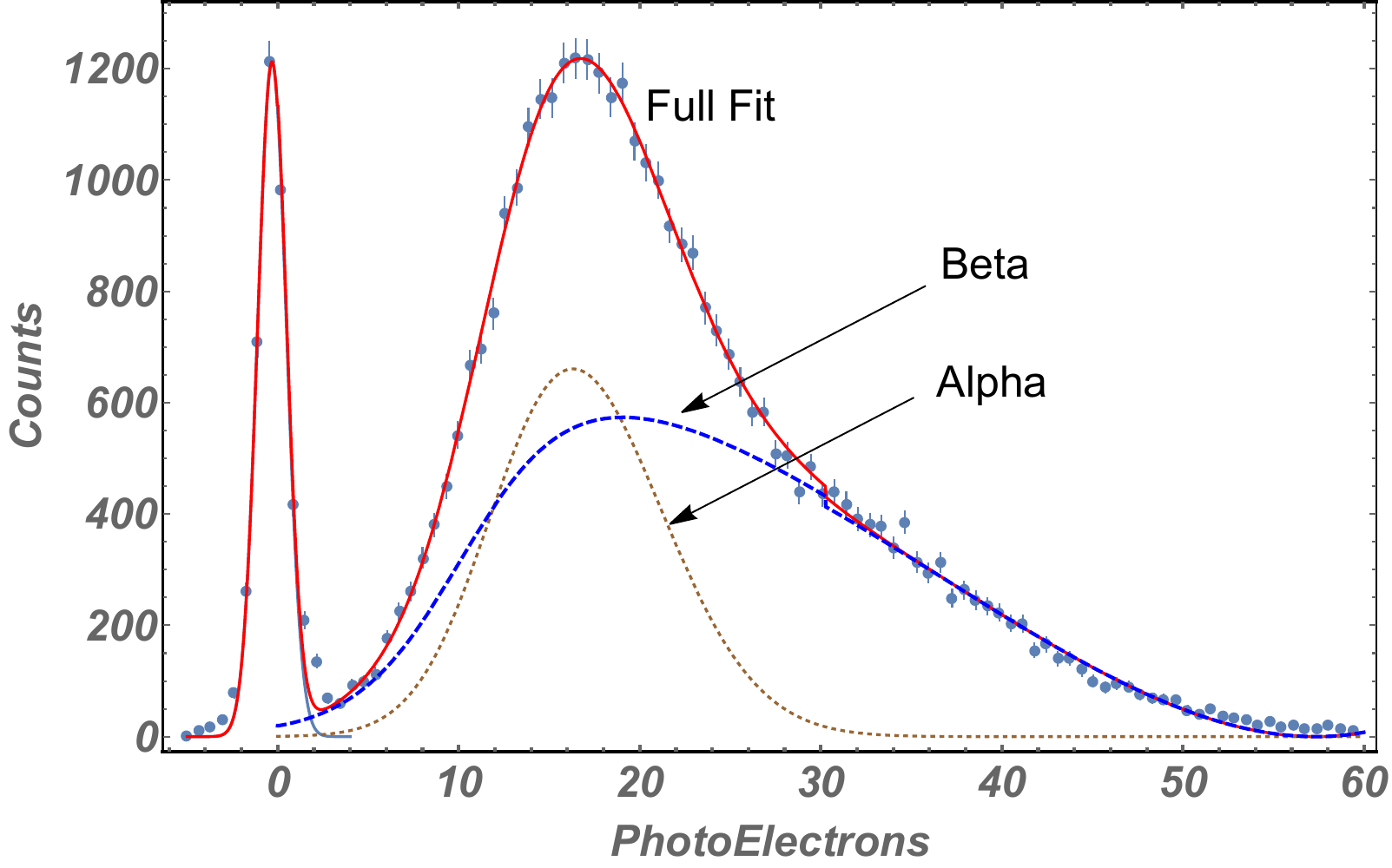}
    \caption{ The total photoelectron yield from a single day of data from the calibration source with a model for $\alpha$ and $\beta$ components. The yield was obtained by summing the signals over all PMTs. The pedestal at zero  is due to accidental triggers. The alpha (peak at 16 pe)  and beta (tail extending to 60 pe) components in the spectrum are separated with a  model that includes a rising threshold at 10 pe. The red curve is the total spectrum including all components, and the blue dashed curve is the beta component.  }
    \label{alpha_pe}
\end{figure}

After we installed this calibration system we were able to perform a number of studies in a continuous manner for several months.  First, all the PMTs on the detector received scintillation light induced by the source. Most of the hits were single photo-electrons 
which could be fitted in the spectrum to obtain the gain 
every few hours for each PMT. This allowed a tight control of the calibration.  The total number of photoelectrons summed from all PMTs is expected to have a combined spectrum of a $\beta$ spectrum with an endpoint of 1161 keV and a broad  Gaussian peak due to the saturated yield from the $\alpha$.  The mean and endpoint of this spectrum are expected to remain constant throughout data-taking after gain calibration. This allows us to check for the stability of the detector independently of the WbLS fill, since none of the light is generated in the WbLS.  Last, each calibration flash comes with a start time recorded from the small PMT; this allows us to calibrate the times of all PMTs relative to the center with excellent precision.    Thorough analysis of water and WbLS data including data from the calibration source and  muons will be presented in a separate publication.  Here we show the total photo-electron yield using a single day of data from the calibration source in Fig.  \ref{alpha_pe} with a model for $\alpha$ and $\beta$ components \cite{betamodel}.



\section{Initial data taking and WbLS Injection}
\label{injection}
The WbLS 1-ton test bed has gone through a commissioning phase and initial data taking.  We  report on the  data taken from this apparatus from August 2022 to November 2022.  During this period, the calibration relied on the LED pulser, and data was obtained with pure water and 1\% WbLS.  The purpose was to establish the method of measurement of the light yield and stability.  During this period, an instability in the high voltage supply was discovered, and the run was terminated to get this problem fixed from the manufacturer.  The data were sufficiently stable to report for this publication. 

\begin{figure}[t]
    \centering
    \includegraphics[width=\textwidth]{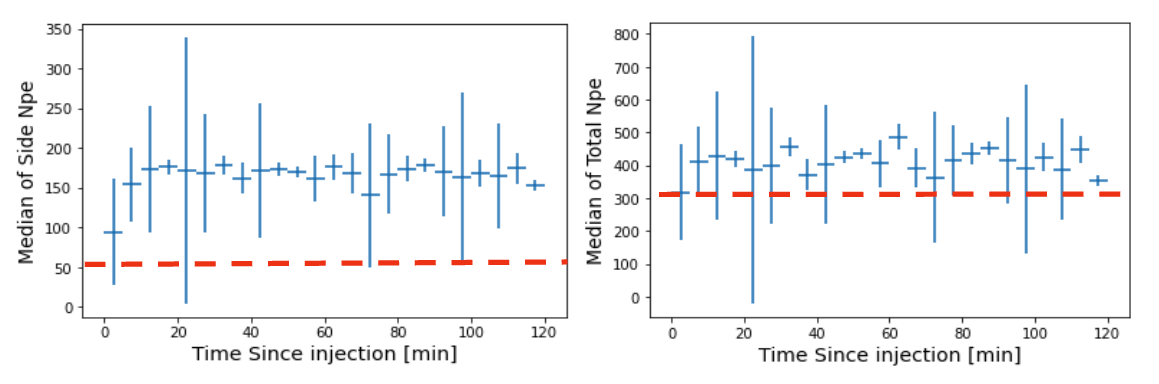} 
    \caption{Left: The median  detected photons in side PMTs from muon tracks. Right: The median detected photons in all PMTs from muon tracks. 
    Each point corresponds to $\sim$5 minutes of data-taking. The large error bars are due to fluctuations in the number of muons.  
    The red dashed line indicates the  light detection level observed during the pure water runs due to Cherenkov light.}
    \label{fig:injection}
\end{figure}

A month-long data taking run with pure water was first obtained.  
All analysis tools and the DAQ system were finely tuned in the pure water run and made ready for the water-based liquid scintillator data. 
We initiated the WbLS phase by introducing 1\% of the organic LS  (by mass) into the 1-ton setup as described in section \ref{sec:organic}. The 1\% concentration was determined by mass measured before and after injection. The instrumentation error on the concentration measurement is less than 3\%. Throughout the injection process, the detector remained online, enabling real-time monitoring of changes in light production from cosmic ray muons. The circulation was also kept running with a  rate of about 6 GPM through a 0.2 micron polypropylene filter.  

Figure \ref{fig:injection} illustrates the variation in detected photons observed in the side and all PMTs after the injection. 
As the organic LS  gradually diffused within the water, the
light yield due to non-Cherenkov light increases. 
This is particularly prominent in the side PMTs since the top two rows of the side PMTs are located above the Cherenkov ring for vertical muons that pass through the trigger counters.  
 In the case of a 1-ton water volume, within just 20 minutes, the liquid scintillator achieved uniform dispersion throughout the tank, as evidenced by the stability of the detected photons. This is a useful outcome 
 of the present R\&D program and 
 underscores the excellent inline mixing of the DIN/PPO organic LS formulation. It should be remarked that the time to complete mixing is lower than the time needed for complete circulation of the tank volume (approximately 50 mins).

\section{WbLS Stability}

The optical stability of the WbLS in our 1-ton detector is assessed using the light yield from bottom and side PMTs as a function of time. 
As mentioned in the previous section, there are water and 1\% WbLS (with a concentration of 15g/L PPO relative to DIN, inside the micelles) phases in the data taking.
We continuously studied the optical properties of the liquid over several weeks using cosmic muons, while keeping the data quality and calibration high and making a few instrumental upgrades along the way.

We employed a sequence of simple data cuts to ensure a robust analysis of light yield (LY) stability. We excluded events flagged for DAQ readout errors or occasional electronic pickup  noise above 1.5 mV. We also excluded rare events in which  the largest pulse peak time fell outside an arbitrarily long time window of 100 ns from the trigger time due to either pile-up or background. We observed an event rate of $(0.74\pm 0.03) \mu$'s/s after the basic quality cuts above. This rate was  constant throughout the data taking period and exhibited an unbiased exponential distribution of times between events, as expected from a Poisson distributed process.  A few 
unstable or poorly calibrated PMTs, identified through continuous gain monitoring, were removed from the analysis. 
The trigger for muon data collection used only the top paddles which simply identified entering muons.  These 
muons could exit either from the bottom or side of the tank leaving a range of energies as well as Cherenkov light topologies in the detector.  
Simple cuts could be performed to select side-exiting (SE) or bottom-exiting (BE) muons.  
A muon that crossed the tank from top and exited from the bottom produced a broad peak which could be easily identified. 


\begin{figure}
    \centering
    \includegraphics[width=0.9\textwidth]{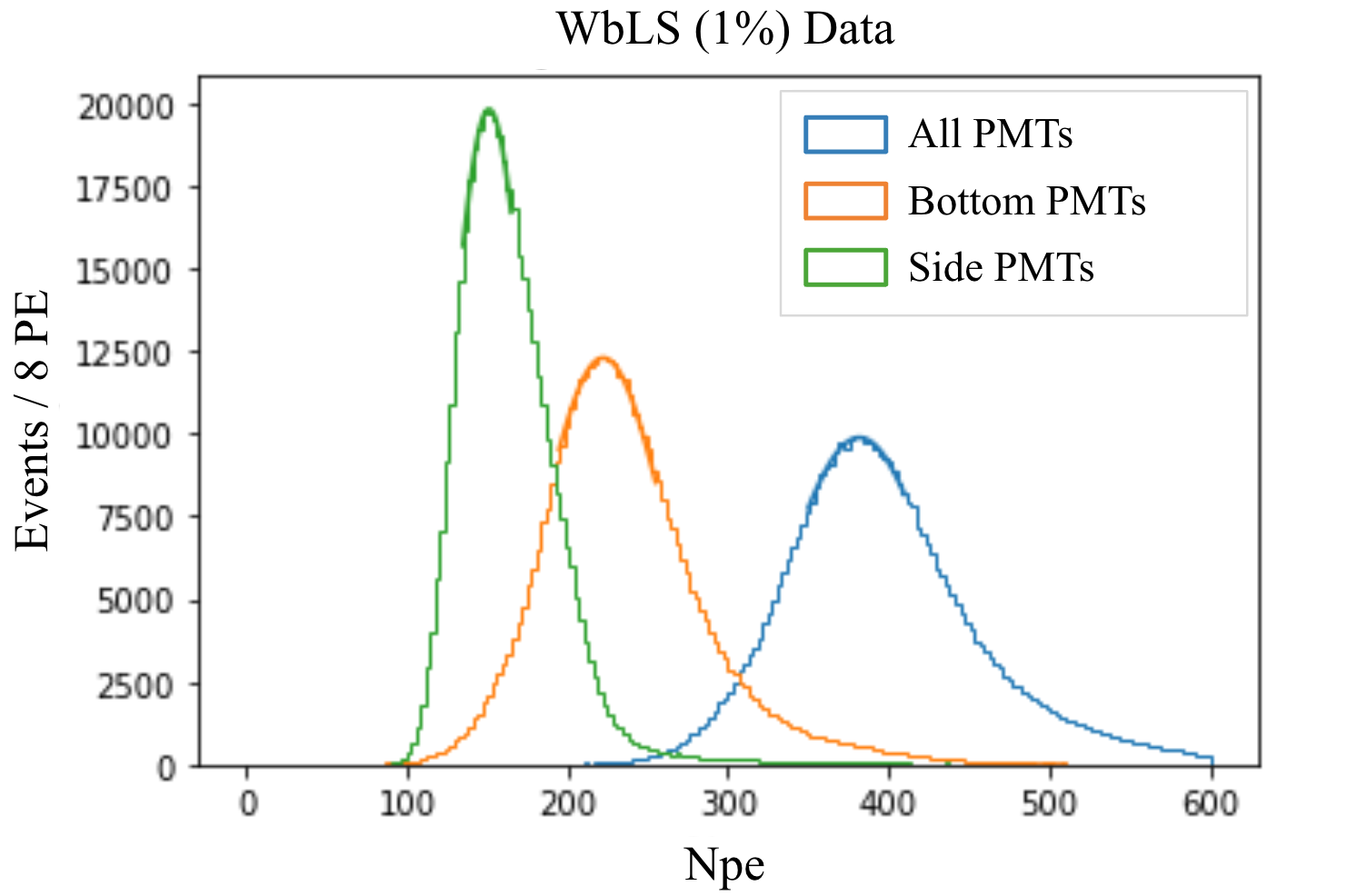}
    \caption{The PE distributions for side, bottom, and all PMTs with a requirement of the sum of photoelectrons for the bottom two rows of PMTs between 65 and 140.}
    \label{fig:pe_peak}
\end{figure}

\begin{figure}
    \centering
    \includegraphics[width=\textwidth]{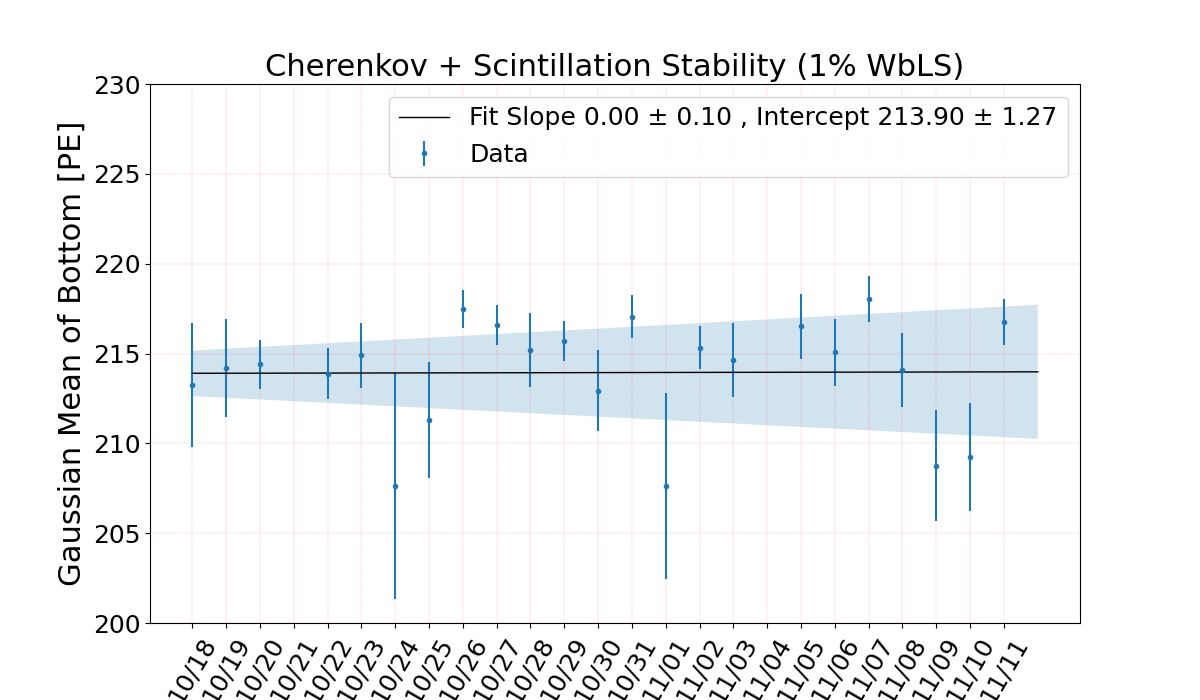}
    \includegraphics[width=\textwidth]{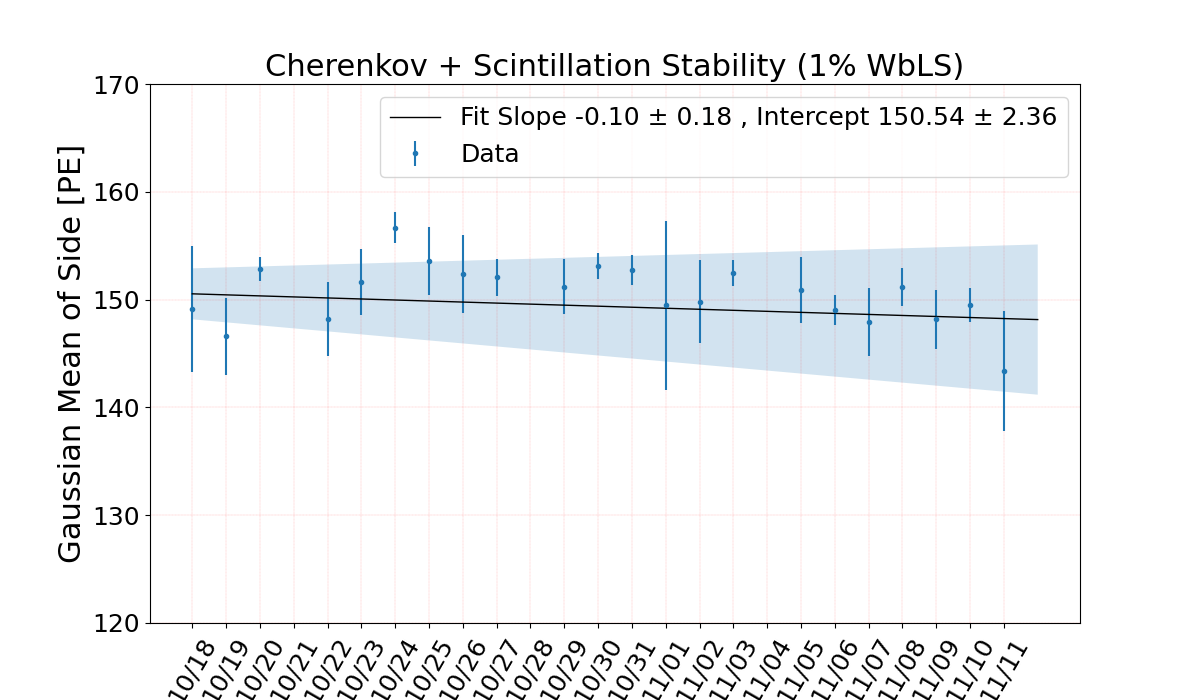}
    \caption{The PE peak stability for groups of PMTs as a function of days. This 25 day period had a stable  LED based gain calibration.  Top: all Bottom PMTs. Bottom: all Side PMTs. The fluctuations are dominated by the muon statistics for that day. The 1-sigma bands are drawn in the plots.
   }
    \label{fig:pe_stability}
\end{figure}


The geometry of the side phototubes was arranged such that the bottom two rows were inside the Cherenkov cone of  a vertical muon, and the top two rows were outside the cone.  
To select BE muons, we summed the number of photo-electrons in the bottom two rows and required the total to be between 65 and 140. 
For the selection criteria, we intended to qualitatively select a Cherenkov-rich and a Cherenkov-poor sample, in order to monitor the stability of the detector with WbLS. More quantitative studies may be required to optimize those criteria.
The photoelectron yield for  side PMTs,   bottom PMTs, and all PMTs for these selected BE muon events is plotted in Fig.~\ref{fig:pe_peak}.  
These data had excellent statistics and the 
 peaks could  be fitted with Gaussian functions and the resulting peak values were used to 
determine the  detector performance stability as a function of time.  
 The result is shown in Fig. \ref{fig:pe_stability}. 
The data starts after 1\% WbLS injection and covers  a period of 25 days when the LED calibration was performed regularly and was stable. 
\textcolor{black}{
The natural timescale for any changes is about 50 minutes which is the time it takes to circulate the tank, but given the continuous nature of filtering, it is unlikely that we can see any such behavior in performance.  We expect the WbLS to be stable both from our experience with laboratory samples and from previous R\&D efforts~\cite{Beriguete:2014gua}. 
} 
To place a limit on any slow variations over days a simple  linear fit was  used to fit for the trend across time; the slope is in  units of PE per day. The stability is excellent within the uncertainty of 0.1 PE per day for the bottom PMTs and 0.2 PE per day for the side PMTs. The side PMTs are expected to have a larger contribution from 1\% WbLS non-Cherenkov light. The small decrease that is fitted in the plot is well within the 1 sigma envelope of the data.  Using the results from the bottom only yield which is dominated by Cherenkov light, we can conclude that the upper bound at 1 sigma is a loss of $< 1.3\%$ of the light per month. Using the side only yield which has a significant non-Cherenkov component, we get an upper bound of $< 3.6\%$ of the light also at 1 sigma. Both of results are however within 1 sigma of no light loss over the course of approximately 1 month. Furthermore there is no visible change in the nature of the WbLS material.  This result is from initial data only; we expect much finer control of both the calibration and the detector stability for data taken concurrently with the calibration source in the future.


\section{Light Yield Analysis}

Detailed analysis of the light yield from muons requires us 
to select well understood muons with trajectories that are restricted to be as vertical as possible and passing through the 
detector.  We therefore used 
 crossing muons (CM) tagged by both the top and bottom paddles. The bottom paddles signals were tagged in readout and resulted in 
 a crossing muon rate of about 1 muon per 2.4 minutes. About 600 muons could be collected every day.  
Since the muons passed through the top and bottom paddles, each with dimension of $\sim$10 cm, the angular variation for these muons was restricted to be $<100 $ mrad and therefore  the length variation was restricted to be $<| 1-(1/\cos(0.1))| = 0.5\%$. This sample of muons, although fewer in number compared to all the bottom exiting (BE) muons, 
provides us with a controlled sample, for a detailed understanding of the light yield.


The muon simulation was carried out using GEANT4 and RATPAC optical propagation software~\cite{ratpac, ratpac2}.  
Muons that passed through the area of the upper and lower scintillator paddles were selected for further analysis.  
The simulation took into account the physics processes of  muon energy deposition, light generation, propagation and light detection. The measured optical properties of the acrylic, optical cookies, and PMT glass window as well as the quantum efficiency of the PMT were used in the simulation model.  
The light generated and then detected in the simulation is recorded in two categories.  
\begin{itemize}
    \item Cherenkov light: the light generated through the Cherenkov process and detected by the PMT directly without  absorption or re-emission.
    \item Non-Cherenkov light: the scintillation light generated from the scintillator or the light that was  re-emitted after absorption. The absorbed light can be either Cherenkov light or scintillation light. 
\end{itemize}
Our data and analysis here concerns cosmic ray muons that are all above Cherenkov threshold. Given that the scintillation and the re-emitted photons have the same geometric, timing, and wavelength characteristics in our model, it is not possible to distinguish them in the current data set using the data alone. 

\begin{figure}
    \centering
    \includegraphics[width=0.9\textwidth]{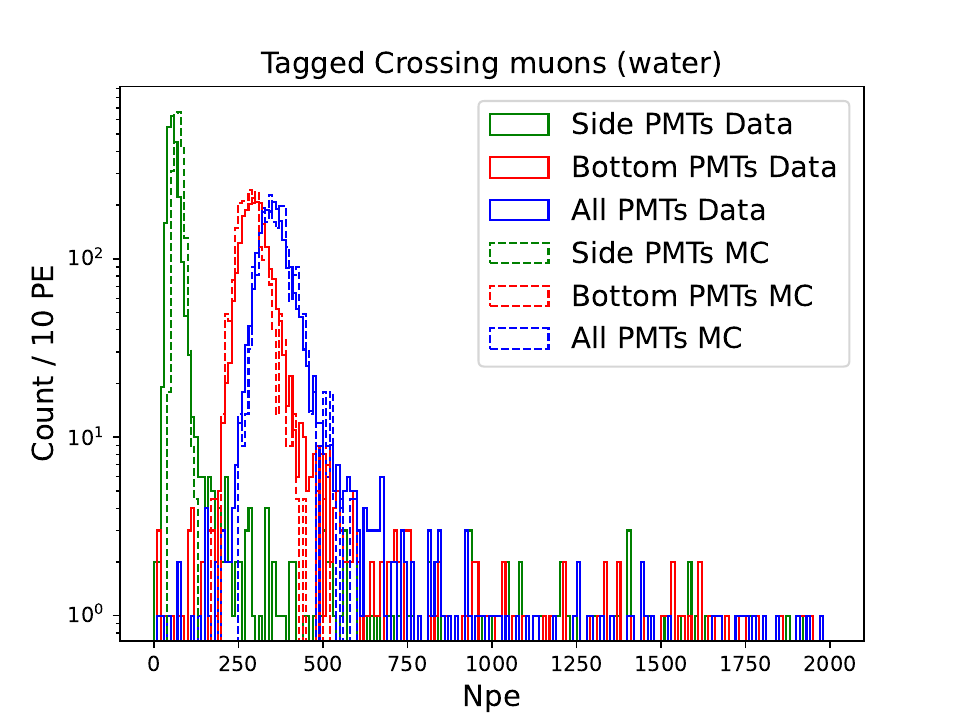}
    \includegraphics[width=0.9\textwidth]{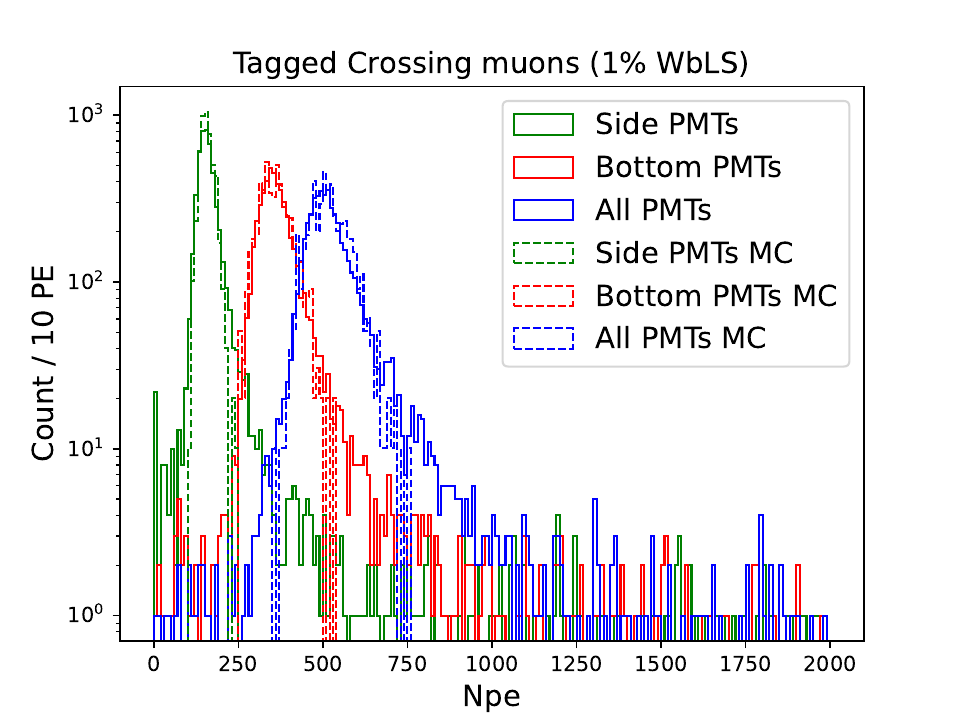}
    \caption{The tagged crossing vertical muons (CM) as observed by the side, bottom and all PMTs. Top: the tank is filled with pure water. The data is solid and simulation is dashed.  
    Bottom: the tank is filled with 1\% WbLS. 
    The data is solid  and simulation is dashed.  
    The simulation for the WbLS plots (bottom) includes the reweighting procedure as described in the text.  
     The additional variance due to PMT efficiencies is also applied to the simulation for both plots  as described in the text.  
     }
    \label{fig:water_muon}
\end{figure}

The muon data and the Monte Carlo for pure water and 1\% WbLS are shown in Fig. ~\ref{fig:water_muon}. 
The photo-electron yield for side and bottom tubes is summed after calibration  and is shown in the figure for the entire period of data (August 2022 to November 2022) to have sufficient statistics for the tagged crossing muons.  Both the mean number of photo-electrons and 
the variance on the distributions are of interest.  The pure water Cherenkov light yield for the bottom (side) is measured to be $297 \pm 37 $ ($56 \pm 13$) photo-electrons. And for the 1\% WbLS data the light yield for bottom (side) is measured to be $350\pm 37$ ($154\pm 22$).  The significant increase on the side is due to the non-Cherenkov component  detected by PMTs outside the Cherenkov cone.  The generation and propagation of the Cherenkov light are well simulated by the Monte Carlo  with little adjustment needed for the detector simulation parameters according to the data-MC comparison for the photo-electron distribution.  For non-Cherenkov light there are two assumptions in the simulation that the non-Cherenkov light is isotropic (by definition), and that it has a spectrum according to the known emission of PPO \cite{buck_2016}. The simulation of Cherenkov, and non-Cherenkov scattering and absorption is modelled in the Monte Carlo. The modeling of the WbLS emission wavelength dependence may contribute a bias to the extracted number of photons per MeV.  This is considered as part of the systematic below.  

The observed variances for both data sets in Fig. \ref{fig:water_muon} have four contribution which we have analyzed: 
\begin{enumerate}
    
\item 
Variations of muon length or trajectories in the detector: As explained above the tagged crossing muons have a negligible variation, nevertheless it is included in the Monte Carlo. 

\item  The variation of the mean muon energy loss or Cherenkov yield:  The muons at sea level have an approximately flat distribution up to 1 GeV and then falls with a power-law distribution.   The energy and angular distribution
are well modelled by the simulation using a reweighting approach~\cite{ParticleDataGroup:2020ssz}. 
The muon energy distribution causes the energy loss to have a distribution that tails off beyond the minimum ionization value for water ($2.03$ MeV/(gm/cm$^2$)).  This distribution has a contribution to  the standard deviation in Fig. \ref{fig:water_muon} of $<4\%$  for the scintillation component. It also contributes to the tails of the distribution at $>3$ sigma which are not well-modelled in the Monte Carlo.  

\item  Variations of geometric issues, collection and quantum efficiencies of the photo-mutlipliers:   The PMT quantum efficiency varies by about 15\% as shown in Fig. \ref{fig:qe}. We assume this variation applies to all the PMTs. In addition,  we expect a larger variation due to the imperfections in the optical coupling of the PMTs and the tank. This variation can be characterized by comparing the expectation of light yield from well modelled Cherenkov light using the pure water data. This variation was found to be random and distributed with a standard deviation of  $\pm$26\% (with a flat distribution) over all PMTs for the bottom.  We assume the same variation for the side PMTs given the PMT coupling for both side and bottom PMTs were done with the same method, and the coupling for the PMTs may change over time due to gravity. A detailed examination will be  in  later publications.
Secondly, we have simulated the acrylic tank as a smooth uniform cylinder with an exact geometry and interfaces in the Monte Carlo. In particular, the total internal reflection of downwards going Cherenkov light from the acrylic-air interface is known to vary  from event to event due to small changes in the angle of incidence on the nonuniform interfaces.
\textcolor{black}{The imperfection of the reflection modeling can be seen in the data versus Monte Carlo distributions 
 for the side PMTs  in the top panel of Fig.~\ref{fig:water_muon}.  }
This incident angle variation combined with the variation of PMT detection efficiencies  results in event by event fluctuation of light yield and an overall widening of the distributions in Fig. \ref{fig:water_muon}  because of  averaging over many PMTs.   
This additional variation  was used as a Gaussian parameter for fitting the Monte Carlo result to the WbLS data as explained below.  
\item  The final and most important contributions to the shape of the photo-electron yield are the energy loss fluctuations and the 
photo-statistics which are well modelled by the Monte Carlo.  

\end{enumerate}


For the 1\% WbLS the Monte Carlo simulation plots in Fig. \ref{fig:water_muon} were obtained by independently 
weighting the two sources of photo-electrons (Cherenkov and non-Cherenkov); no constraint from the pure water data was included.  
The simulation was performed with a physics model for Cherenkov light in pure water and a nominal yield with the expected scintillation spectrum for the non-Cherenkov light,  and then a fit was performed to obtain agreement with the side and bottom yields. The fit had three parameters: the Cherenkov and non-Cherenkov yields, and in addition an overall Gaussian standard deviation  to account for the variance of the PMT collection and quantum efficiencies.  With the three floating parameters, the simulated PE distributions for the side and bottom PMTs are compared with the data in Fig.  \ref{fig:water_muon}. For the 1\% WbLS, the best-fit parameters were extracted by minimizing the difference between the data and simulated distributions. The resulting Cherenkov and non-Cherenkov weights correspond to a mean reduction of Cherenkov yield by 5\% and a non-Cherenkov yield of 
127.6 $\pm$ 17.6
photons per MeV. The Cherenkov weight for the pure water fit indicates a mean reduction of Cherenkov yield by only 2\%. 
The ratio of Cherenkov to non-Cherenkov yields is detector dependent, given the sensitivity as a function of wavelength, but for our setup it is  well measured to be $1.9 \pm  0.3 $. 
The fitted additional variation due to geometric factors and PMT efficiencies corresponds to a 7.5\% (1 sigma)  increase to the muon yield variation.  All numbers quoted above are with statistics only 1 sigma errors from the fits.  



We expect  the systematic uncertainty on the light yield analysis to be mainly from  the PMT efficiency and geometric performance including reflections and optical couplings to the detector. 
To evaluate any potential  bias,  we randomly assigned the efficiencies to all PMTs following the measured distribution from the bottom PMTs with water only Cherenkov data. 
A new photo-electron distribution for Monte Carlo was obtained for the randomized efficiencies, and compared to data to extract the yields using the weighting method.  
As an example, for a particular trial resulting  weights were found to correspond to overall weight of 0.96 for Cherenkov light and a yield of 55 photons per MeV for non-Cherenkov light. 
After  100 repeated trials, each with randomly assigned efficiencies drawn from the measured distribution,  the standard deviation is taken as the systematic uncertainty for the mean light yield value.
As remarked earlier, the wavelength dependence of the PPO emission combined with absorption in the fluid, the acrylic walls, optical couplings, and quantum efficiency is included in the Monte Carlo, and it is used for extraction of the non-Cherenkov light. Most of this dependence comes from the PPO emission coupled to the photomultiplier quantum efficiency. This systematic error is examined by varying the PPO emission within 10 nm (as suggested in \cite{buck_2016, yeh2024gd}), and the systematic is found to be $<5\%$ in the non-Cherenkov light yield because the PPO emission is rather well matched to the various optical windows and the PMTs.  This is added in quadrature to the main systematic from PMT and geometric efficiency effects.

This study concludes that the  light yield for the 1\% WbLS material using DIN as the main organic scintillator  
produces Cherenkov light with a weight factor of 0.95 $\pm$ 0.08 (syst.) $\pm$ 0.11 (stat.) with respect to the nominal Geant4.11 Cherenkov yield for pure water; this is consistent with no loss of Cherenkov yield at 1 sigma.  The result for 
 non-Cherenkov light is a  light yield of 
 127.6 $\pm$ 19.8 (syst.) $\pm$ 17.6 (stat.)
 photons per MeV of ionization deposit for minimum ionizing particles. Comparison of this measurement with benchtop measurements is in progress\cite{yeh2024gd}. 
 
 
In addition, the same WbLS was also deployed at ANNIE/SANDI at FNAL~\cite{ANNIE:2023yny}. No deterioration of optical transparency or light-yield has been found, indicating the WbLS stability under transportation, deployment, and beam measurement conditions over 6 months of operation. 
An increase in the detected number of photoelectrons in WbLS-filled SANDI was observed for through-going muons, Michel electrons from muon decays, and beam neutrino interactions compared to water. 
However, similar to the results reported in this paper, the scintillation light yield measured at SANDI also requires detailed Monte Carlo simulation including a model of the full ANNIE detector response with the SANDI vessel structure and light propagation mechanism traversing through the WbLS liquids.
 A detailed Monte Carlo model to account for the light propagation mechanism including absorption, re-emission, and scattering of WbLS in the full 1-ton detector vessel is currently underway. Further work is needed to confidently separate the components of the non-Cherenkov light to obtain the intrinsic scintillation light-yield of the WbLS.  



\begin{figure}
    \centering
    \includegraphics[width=0.9\textwidth]{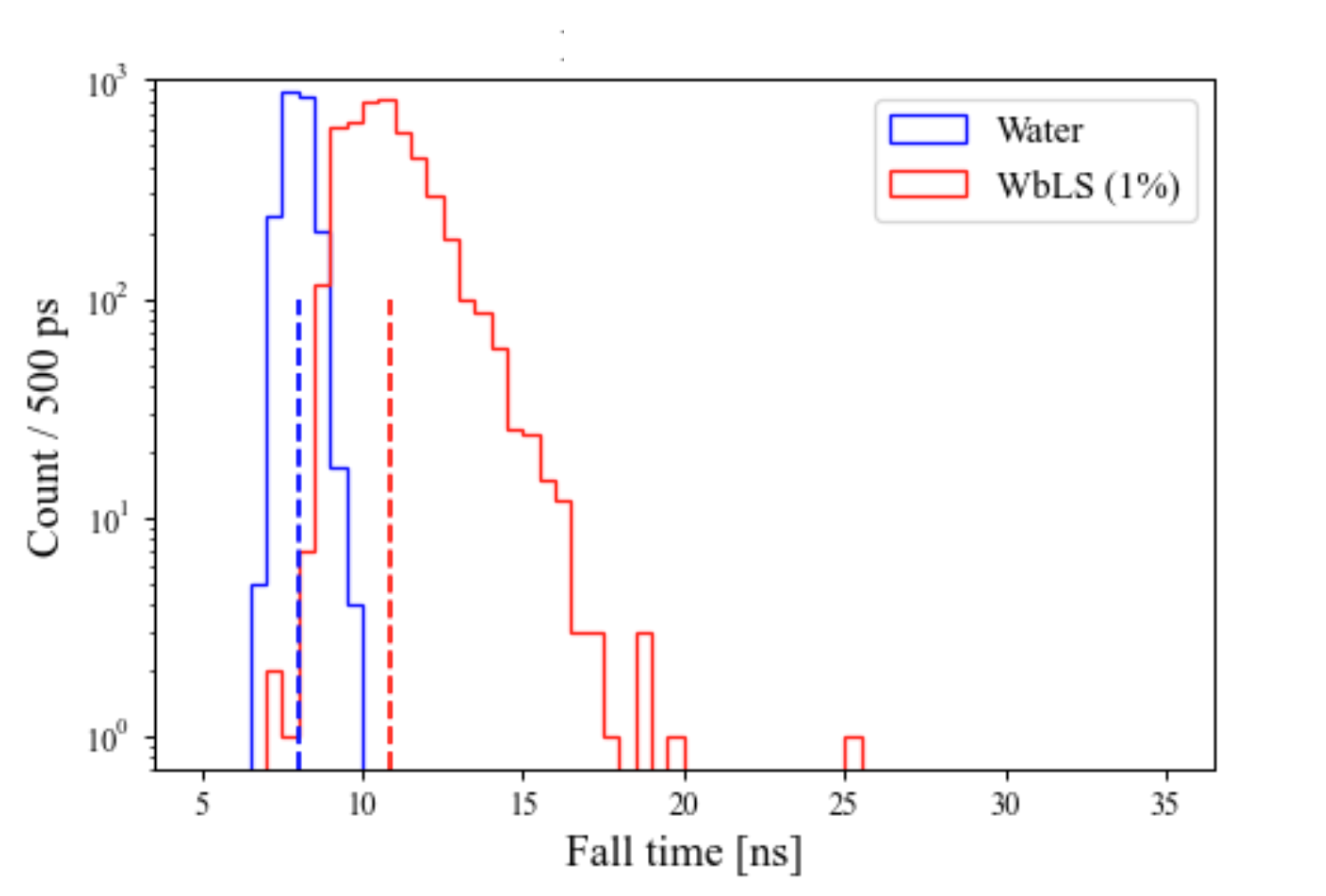}
    \caption{A comparison of pulse-shape differences between water and 1\% WbLS data for the tagged crossing muons (CM) is shown. The signals are observed in the bottom PMTs. The fall time is defined as the time required to decrease from 90 to 10 percent of the pulse height. Dashed lines are the means.
     }
    \label{fig:fall_time}
\end{figure}

Lastly, the tagged crossing muons show a distinct pulse shape difference in the tail region due to the slower re-emission or scintillation component, as shown in Fig~\ref{fig:fall_time}. This preliminary result indicates that analyzing the Cherenkov-to-scintillation ratio (C/S) in hybrid neutrino detectors could enable effective pulse-shape discrimination for particle identification with this new WbLS formulation. Moreover, the DIN formulation allows for the adjustment of the scintillation's fast-to-slow component ratio, making it a versatile target material for the next-generation neutrino detectors. A comprehensive analysis of C/S separation will be presented in future work. 

\textcolor{black}{In the near future, we plan to reduce both the statistical and systematic uncertainties on the measured light yield with new data from with stable detector conditions. Muons will be tracked with a new pair of hodoscopes to constrain the geometry.  Data will be taken in several configurations, and also with smaller increments of WbLS filling to tune the optical and scattering model. From laboratory samples, we have seen no difficulties in concentrations ranging from very small to several percent regarding dissolution, uniformity, or stability. 
We also expect to learn a lot more concerning the optical model with the  4ton Eos detector at Berkeley~\cite{Anderson:2022lbb} and the 30ton demonstrator at BNL that is  being commissioned.}


\section{Conclusion} 

We have constructed and operated a 1-ton scale Water-based Liquid Scintillator detector tank. The details of the instrumentation and the initial data are described above in detail. The detector was operated with a new cocktail of WbLS based on mixing organic scintillator based on DIN  (di-isopropylnaphthalene), and it is compatible with loading such a detector with metals such as Gadolinium.  
The initial results indicate stability better than a few percent  per month.   With a mixture of 1\% organic scintillator in water, the total light yield (Cherenkov and non-Cherenkov) is significantly enhanced. The non-Cherenkov light yield is measured to be \textcolor{black}{
$127.6 \pm 19.8 (syst.) \pm 17.6 (stat.) $}
photons per MeV for muons.  The   primary source of known systematic uncertainties is due to
our understanding of the optical couplings and efficiencies of the photomultipliers. The errors are  considerably mitigated by the statistics of having a large number of PMTs and also the geometry of the apparatus.  
However,  other potential uncertainties due to absorption of non-Cherenkov light based on the  PPO emission spectrum  are not yet ruled out. 
The extraction of the light yield depends on the assumption that  both components of the non-Cherenkov light -- the scintillation light and the re-emitted light --  have the same PPO emission spectrum and are isotropic. 
A detailed Monte 
Carlo model is needed to examine the light propagation mechanisms including absorption, re-emission, and
scattering to examine the result under variations of the above assumption and for different sample geometries.  
The ratio of Cherenkov to non-Cherenkov yield
is detector and geometry dependent; for our setup it is well characterized to be \textcolor{black}{$1.9\pm 0.4$} including systematics. 

The detector and its fluid handling system are fully operational. Data sets with varying levels of organics (and with potentially different chemical compositions) and with continuous calibration using the radioactive source described above will be provided in the near future.

\acknowledgments


The work conducted at Brookhaven National Laboratory was supported by the U.S. Department of Energy under contract DE-AC02-98CH10886. 
Work conducted at Lawrence Berkeley National Laboratory was performed under the auspices of the U.S. Department of Energy under Contract DE-AC02-05CH11231.  The project was funded by the U.S. Department of Energy, National Nuclear Security Administration, Office of Defense Nuclear Non-proliferation Research and Development (DNN R\&D).  
This material is based upon work supported by the U.S. Department of Energy, Office of Science, Office of High Energy Physics, under Award Numbers DE-SC0018974, DE-SC0012704 and DE-SC0012447.





\bibliographystyle{unsrt}
\bibliography{biblio} 


\end{document}